\documentclass[twocolumn,a4paper]{aa}
\usepackage{epsf,times}

\begin{document}

\thesaurus{03(11.09.4; 11.01.1; 11.05.2; 09.01.1; 11.19.2)}

\title{Star formation and the interstellar medium in low surface
brightness galaxies}

\subtitle{I.  Oxygen abundances and abundance gradients in low surface
brightness disk galaxies}

\author{W.J.G.  de Blok\thanks{\emph{Present address:} School of
Physics, University of Melbourne, Parkville VIC 3052, Australia} and
J.M.  van der Hulst}

\institute{Kapteyn Astronomical Institute, P.O.~Box 800, 9700 AV
  Groningen, The Netherlands,\\ edeblok@physics.unimelb.edu.au,
  vdhulst@astro.rug.nl }
\offprints{W.J.G. de Blok}

\date{Received: ; accepted: }
\maketitle

\markboth{de Blok \& van der Hulst: Star formation \& ISM in LSB
  galaxies}{}

\hoffset=5mm
\voffset=10mm
\def\degr{$^\circ$}
\def\magn{mag arcsec$^{-2}$}
\def\r23{\hbox{R}$_{23}$}
\def\o32{\hbox{O}$_{32}$}
\def\logoh{$\log(\hbox{O}/\hbox{H})$\ }
\def\<U>{$\langle U \rangle$}

\begin{abstract}
  We present measurements of the oxygen abundances in 64 H{\sc ii}
  regions in 12 LSB galaxies. We find that oxygen abundances are
  low. No regions with solar abundance have been found, and most have
  oxygen abundances $\sim 0.5$ to 0.1 solar.  
  The oxygen abundance appears to be constant as a function of radius,
  supporting the picture of quiescently and sporadically evolving LSB
  galaxies.
\end{abstract}
\keywords{Galaxies: ISM -- Galaxies: abundances -- Galaxies: evolution
  -- ISM: abundances -- Galaxies: spiral}

\section{Introduction}

Low surface brightness (LSB) disk galaxies have all the
characteristics of unevolved galaxies. Those discovered so far
constitute a population of gas-rich, metal-poor galaxies with very low
star formation rates (see the review by Bothun et al.
1997).  Their surface brightnesses are a few magnitudes lower than the
values commonly found for so-called normal galaxies.  Most of them are
rather late-type galaxies, with diffuse spiral arms.

A direct probe of the evolutionary state of these galaxies is the
metal abundance in the interstellar medium (ISM).  A low abundance
generally indicates only limited enrichment of the ISM and therefore
(in a closed system) a small amount of evolution.

Because of their low surface brightness, obtaining spectra of the
stellar disks of LSB galaxies is difficult and requires large amounts
of telescope time.  Conclusions on metallicities must therefore be
derived from spectra of H{\sc ii} regions.  These usually are the brightest
distinct objects in a LSB galaxy.  Their bright emission lines make
them more easily observable than the underlying continuum.  The H{\sc ii}
regions that are observed in LSB galaxies are usually giant H{\sc ii}
regions, that are ionized by star clusters rather than by a few stars.

The first measurements of the oxygen abundances in H{\sc ii} regions in LSB
galaxies were presented in McGaugh (1994). He found, using an
empirical oxygen abundance indicator, that LSB galaxies are low
metallicity galaxies with typical values for the metallicity
$Z<\frac{1}{3} Z_{\odot}$.  It shows that low metallicities can occur
in galaxies that are comparable in size and mass to the bright
galaxies that define the Hubble sequence. As LSB galaxies are found to
be isolated (Mo et al.\ 1994), this suggest that surface mass density
and environment are as important for the evolution of a galaxy as
total mass.

In this paper we present a follow-up study of oxygen abundances in LSB
galaxies. We confirm the results by McGaugh (1994) that LSB galaxies
are metal-poor. We present two direct measurements of the oxygen
abundance from measurements of the [O {\sc iii}]$\lambda 4363$ line,
supplemented with a large number of empirically determined oxygen
abundances.  In addition, for those galaxies where sufficient data are
available, we investigate the change in abundance with radius, and
show that the measurements are consistent with no gradient. The
steeper gradients found in HSB galaxies (Vila-Costas \& Edmunds 1992
[VCE], Zaritsky et al. 1994, Henry \& Howard 1995, Kennicutt \&
Garnett 1996) are not present.  It is worth noting that the exact
form and magnitude of the Milky Way oxygen gradient has now been
consistently reproduced in early-type stars, H{\sc ii} regions and
planetary nebulae, which supports that the extra-galactic H{\sc ii}
region gradients in HSB galaxies are real (Smartt \& Rolleston 1997).

The lack of abundance gradients in LSB galaxies supports the picture
of stochastic and sporadic evolution, where the evolutionary rate only
depends on local conditions and not on the global properties of LSB
galaxies as a whole.

Section 2 describes the sample selection and observations. Section 3
presents the data, while in Sect.~4 the analysis is
described. Section 5 discusses the abundances found. Section 6
discusses reddening towards the H{\sc ii} regions, while Sect.~7
concludes with presenting the gradients. In Sect.~8 the results are
summarized.

\section{Sample selection, observations and reduction}

LSB galaxies from the samples of van der Hulst et al.\ (1993) and de
Blok et al. (1995) were examined for the presence of
H{\sc ii} regions using H$\alpha$ images.  We selected as possible targets
only those galaxies with two or more distinct H{\sc ii} regions.  This
introduces a bias in our sample, and it therefore cannot be considered
complete in any sense. Nevertheless it makes it possible to study the
ISM in a significant subset of the LSB galaxy population.

Optical spectra of 64 H{\sc ii} regions in 12 of the selected LSB galaxies
were obtained from 6--10 February 1992 with the 4.3 meter William
Herschel Telescope (WHT) at La Palma.  The ISIS spectrograph was used.
The gratings were adjusted such that the blue arm covered a wavelength
range from $\sim 3600$ Angstrom  to $\sim 5400$ Angstrom, while the red arm
covered a range from $\sim 5100$ Angstrom to $\sim 6800$ Angstrom.  An amount
of overlap therefore existed, which was used to tie the spectra
together. The resolution for both arms was 1.4 Angstrom per pixel in the
dispersion direction.

Conditions during the observing run were variable: usually photometric
but with seeing varying between 2\arcsec\ and 5\arcsec.  Total exposure
times per long-slit spectrum were 4000 seconds.  A slit width of 1\arcsec\
was used during the first two nights.  This was later adjusted to
1.6\arcsec\ for the last two nights.

\begin{table*}
\begin{minipage}{110 mm}
\caption[]{Balmer line data}
\begin{tabular}{lcrrrrrrr}
\hline
{Galaxy}& {H{\sc ii}}& {$F(H\beta)$} & {$\sigma_{H\beta}$}& 
{$W_{\lambda}$}&{$W_{\lambda}$}&
{$W_{\lambda}$}&$c$&$\sigma_{c}$\\
& & & &(H$\gamma)$&(H$\beta)$&(H$\alpha$)& &\\
\hline
F561-1&        1-1&   197&      9&        &   30&   112 & 0.054 & 0.057   \\
      &        1-2&   574&      14&        &   47&   107 & 0.141 & 0.029  \\
F563-1&        2-1&   188&      7&      43&  241&   187 & --0.397\rlap{::} & 0.080   \\
      &        2-2&   139&      7&        &   43&   102 & --0.461\rlap{::} & 0.069   \\
      & \llap{*}2-3&  1278&      20&      54&  114&   110 & 0.087 & 0.018  \\
       &        2-4&    92&      6&        &   19&   103 & 0.709 & 0.087  \\
       &        4-1&   162&      6&        &   20&   120 & 0.335 & 0.049  \\
       &        4-2&  1105&      16&      10&   31&    62 & 0.321 & 0.017 \\
       &        4-3&    71&      5&        &    8&    68 & 1.347 & 0.093  \\
F563-V1&        3-2&   151&      7&        &   21&    52 & --0.072 & 0.074  \\
       &        3-3&   104&      7&        &   14&    26 & --0.002  & 0.089  \\
F563-V2&        2-1&   359&      10&      15&   52&    22 & --0.174\rlap{:} & 0.037  \\
       &        2-2&   588&      14&      28&   74&   239 & 0.191 & 0.029  \\
       &        2-3&   232&      9&      16&   26&   136 & 0.277  & 0.047  \\
       &        4-1&   409&      11&        &   14&    92 & 0.519 & 0.031  \\
       &        4-2&   240&     10&        &    6&    31 & 0.441  & 0.048 \\
F568-3 &        2-1&   193&      8&      20&   54&   145 & 0.250  & 0.050 \\
F568-V1&        3-1&   103&      7&        &    6&    39 & 0.503  & 0.086 \\
F571-5 & \llap{*}3-1&   476&      11&      25&  272&   351 & --0.202\rlap{:} & 0.028 \\
       & \llap{*}3-2&  1266&      17&      28&   80&   272 & --0.038 & 0.016 \\
F571-V1&        3-1&    86&      4&        &   37&    85 & 0.370 & 0.086 \\
U1230  &        1-1&   196&     7&        &   16&   323 & 0.072 & 0.054 \\
       &        3-1&   163&      7&        &   14&    39 & 0.162 & 0.064 \\
       &        3-2&   145&      7&        &   43&    20 & --0.435\rlap{::} & 0.110  \\
       &        3-6&   139&      7&        &   52&  2615 & 0.645 & 0.063 \\
U5005  &        2-1&   156&      7&        &   49&   117 & 0.129 & 0.067 \\
       & \llap{*}2-2&  1089&      18&      43&  123&   501 & 0.249 & 0.021 \\
       &        2-3&    88&      6&      93&   51&    67 & --0.120\rlap{:} &0.104 \\
       &        2-4&   139&      7&        &   45&   145 & 0.415 &0.069 \\
       &        2-5&   103&      6&        &   18&    59 & 0.120  & 0.079 \\
       &        2-6&    86&      6&        &   19&   143 & 0.243 & 0.087 \\
       &        2-7&    46&      5&        &   15&    47 & 0.105 & 0.143 \\
       &        3-2&   162&      6&        &   20&    57 & --0.542\rlap{::} & 0.063 \\
       &        3-3&   184&      10&      22&   42&    87 & 0.194 & 0.065 \\
       &        3-5&   154&      7&       &   29&   122 & --0.003 & 0.078 \\
       &        4-1&   610&     13&        &   12&    51 & 0.442 & 0.026 \\
       &        4-2&   163&      7&        &   13&    60 & 0.887  & 0.052 \\
U5999  &        3-1&   185&      8&        &   14&   850 & 0.324 & 0.057 \\
       &        3-2&   235&      10&        &    9&    60 & 0.488  & 0.050 \\
       &        3-3&   233&      8&        &   17&   103 & 0.170 & 0.047 \\
       &        3-4&   540&     13&        &   14&    52 & 0.142 & 0.028 \\
       &        4-1&   163&      8&        &    8&   114 & --0.058 & 0.079 \\
       &        4-4&   374&      10&        &   23&    49 & 0.458 & 0.034 \\
\hline
\end{tabular}
\end{minipage}
\end{table*}

\begin{table*}
\begin{minipage}{110 mm}
\caption[]{Line ratios}
\begin{tabular}{lcrrrrrrrr}
\hline
{  Galaxy}& {  H{\sc ii}}& {  [O {\sc ii}]} & {  H$\gamma$}& 
{  [O {\sc iii}]}&{  [O {\sc iii}]}&
{  H$\alpha$}&{  [N {\sc ii}]}&{  [S {\sc ii}]}&{  [S {\sc ii}]}\\
&&$\lambda 3727$&$\lambda 4861$&$\lambda 4959$&$\lambda 5007$&$\lambda
6563$& $\lambda 6583$&$\lambda6717$&$\lambda 6731$\\
\hline

F561-1&1-1&2.566&&0.447&1.836&2.994&0.524&0.595&0.350\\
 &1-2&2.862&&0.725&1.948&3.224&0.169&0.421&0.249\\
F563-1&2-1&3.158&0.552&0.556&1.472&2.040&0.130&0.328&0.209\\
 &(2-2)&3.383&&0.425&1.688&1.931&0.199&0.489&0.277\\
 &\llap{*}2-3&1.232&0.417&1.755&5.118&3.079&0.088&0.171&0.134\\
 &(2-4)&4.733&&0.652&2.501&5.234&0.417&0.926&0.543\\
 &4-1&3.465&&0.451&1.538&3.805&0.226&0.673&0.331\\
 &4-2&1.636&0.270&1.586&4.646&3.758&0.114&0.294&0.202\\
 &(4-3)&5.003&&0.592&2.270&9.007&0.513&2.334&1.240\\
F563-V1&3-2&3.745&&1.418&3.704&2.690&&0.362&0.246\\
 &(3-3)&3.522&&0.826&2.647&2.855&&0.596&0.541\\
F563-V2&2-1&1.748&0.316&0.916&2.105&2.465&0.302&0.293&0.268\\
 &2-2&1.460&0.328&1.181&3.211&3.366&0.150&0.347&0.273\\
 &2-3&3.754&0.468&0.887&2.660&3.620&0.249&0.466&0.321\\
 &4-1&4.515&&0.942&3.027&4.450&0.285&0.973&0.897\\
 &4-2&5.058&&0.754&1.779&4.163&0.505&0.909&1.296\\
F568-3&2-1&2.955&0.563&0.726&1.743&3.540&0.489&0.456&\\
F568-V1&(3-1)&6.641&&1.761&1.582&4.391&0.392&0.898&1.234\\
F571-5&\llap{*}3-1&1.460&0.275&1.539&4.467&2.408&0.827\rlap{::}&0.140&0.339\\
 &\llap{*}3-2&2.460&0.404&1.194&3.566&2.770&0.144&0.225&0.224\\
F571-V1&(3-1)&2.967&&0.639&2.133&3.919&&&\\
U1230 &1-1&1.557&&0.662&1.464&3.041&0.247&0.270&0.294\\
 &3-1&&&0.871&2.050&3.283&0.282&&\\
 &(3-2)&&&1.511&2.343&1.975&&0.308&0.419\\
 &(3-6)&&&0.718&2.531&4.955&0.568&1.619&1.010\\
U5005 &2-1&2.621&&0.361&0.964&3.192&&0.302&0.450\\
 &\llap{*}2-2&2.577&0.424&1.130&3.333&3.535&0.206&0.294&0.250\\
 &(2-3)&3.165&0.927&1.095&3.404&2.583&&&\\
 &2-4&4.652&&0.653&1.441&4.074&0.803&0.688&0.523\\
 &2-5&&&0.389&1.662&3.168&0.571&0.503&0.683\\
 &(2-6)&4.483&&0.830&3.066&3.519&0.148&0.488&0.513\\
 &(2-7)&4.363&&2.251&4.053&3.127&0.195&&\\
 &3-2&2.526&&0.594&1.548&1.803&0.133&0.397&0.357\\
 &3-3&3.073&0.471&0.603&2.050&3.374&0.324&0.620&0.432\\
 &3-5&3.032&&0.388&1.768&2.854&0.233&0.553&0.328\\
 &4-1&3.618&&0.717&1.874&4.169&0.511&0.525&0.405\\
 &(4-2)&5.167&&1.140&3.761&6.087&0.411&1.103&0.618\\
U5999&3-1&5.281&&0.887&2.554&3.770&0.218&0.702&0.379\\
 &3-2&4.788&&0.566&1.779&4.333&0.686&1.105&0.671\\
 &3-3&4.418&&0.859&2.684&3.305&0.108&0.571&0.479\\
 &3-4&3.316&&0.504&1.804&3.227&0.159&0.383&0.212\\
 &4-1&5.365&&1.144&1.322&2.723&0.280&0.497&0.872\\
 &4-4&3.190&&0.511&1.736&4.224&0.822&0.311&0.291\\
\hline

{  Galaxy}& { \ H{\sc ii}}& {  [O {\sc iii}]}&H$\delta$&{  H$\epsilon$}&
H$\eta$&{  [Ne {\sc iii}]}&He I&\\
& &$\lambda 4363$ &$\lambda 4101$ &$\lambda 3970$ & $\lambda 3889$
&$\lambda 3869$ & $\lambda 5876$& \\
\hline
F563-1   &      2-3&   0.051&   0.191&   0.189&   0.092&   0.371&   0.132&&\cr
F571-5   &      3-1&        &        &        &        &   0.382&   0.134&&\cr
        &      3-2&       &   0.317&   0.118&  0.171&  0.206&  0.091&&\cr
U5005   &      2-2& 0.033& 0.291&  0.204& 0.144& 0.343&  0.129&&\cr
\hline
\end{tabular}
Note: H{\sc ii} regions with their identification number between brackets
are not used for further analysis (see Sect.~5).\\
\end{minipage}
\end{table*}

\begin{figure*}
\epsfxsize=0.5\hsize
\centerline{\hbox{\epsfbox[17 80 580 755]{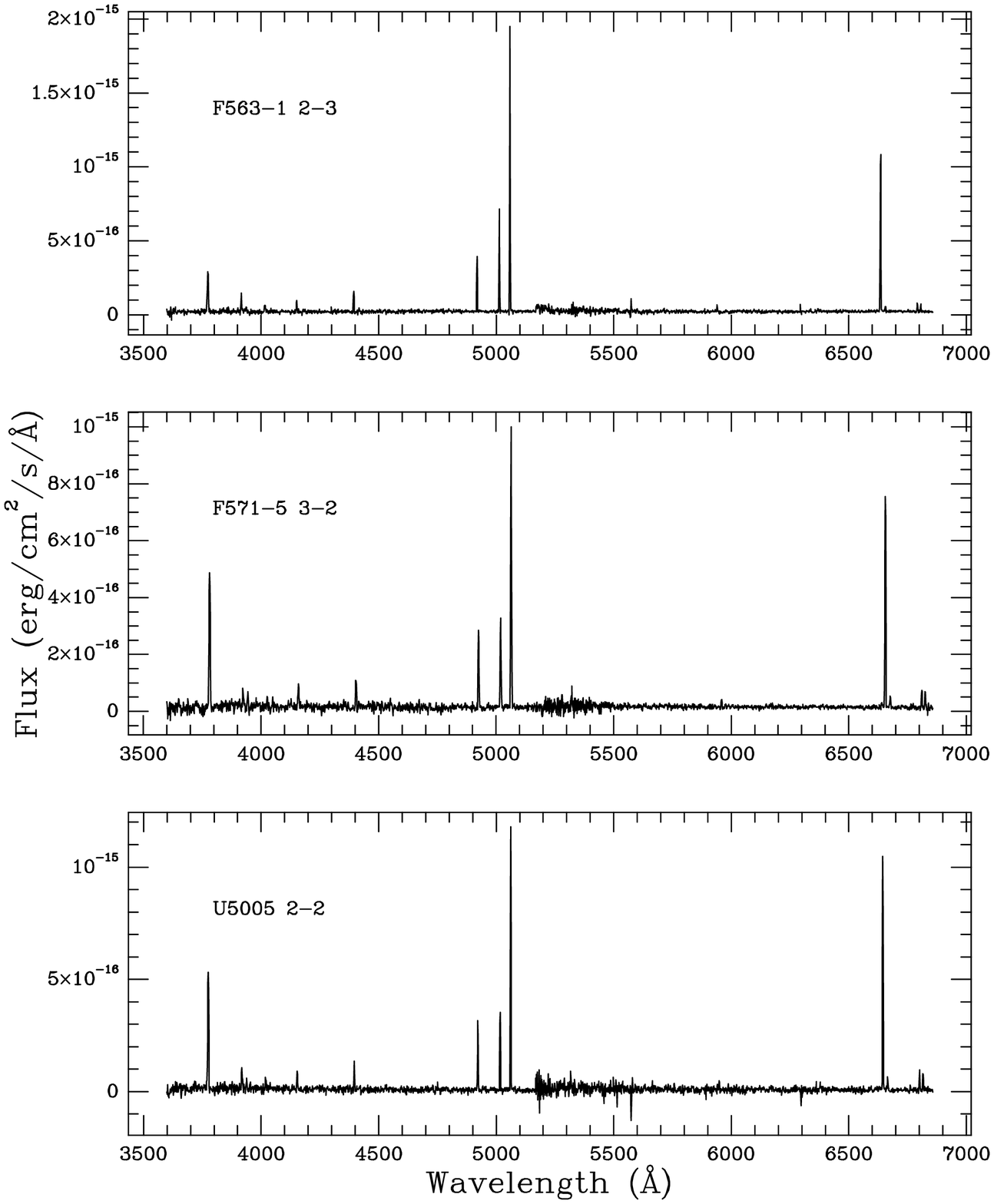}}}
\epsfxsize=0.5\hsize
\centerline{\hbox{\epsfbox[17 273 580 755]{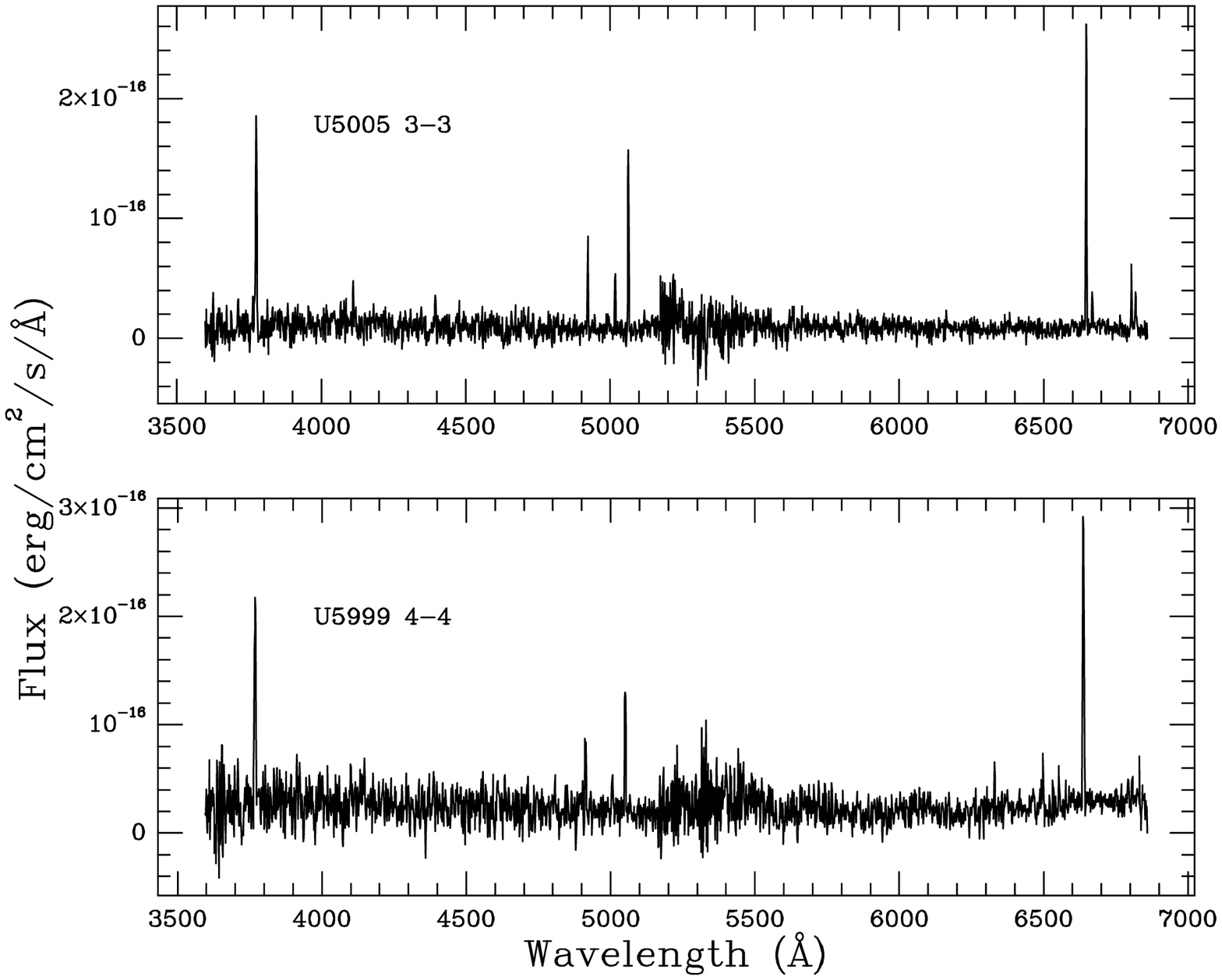}}}
\caption{Some examples of typical spectra of LSB galaxy H{\sc ii} regions.}
\end{figure*}

The data were reduced and analyzed using standard long-slit reduction
tasks from the {\sc iraf} package.  Two independent reductions of the
data resulted in a mean difference between integrated fluxes of the
same H{\sc ii} regions of $\sim 10\%$.  These systematic differences
are not taken into account in further analyses, but should be kept in
mind.

\bigskip
\section{The Data}
Figure 1 shows a few examples of LSB galaxy H{\sc ii} region spectra.
The resulting data (uncorrected for foreground Galactic extinction
which is in most cases less than 0.05 mag in $B$) are presented in
Tables 1 and 2.

Table~1 contains data on the hydrogen Balmer emission lines. Column 1
gives the name of the galaxy.  Column 2 contains the H{\sc ii} region
identification.  The first number refers to the night the spectrum was
taken; the second number is an arbitrary ranking number based on the
position of the H{\sc ii} region along the slit. 

Column 3 contains the H$\beta$ flux in units of $10^{-18}$ erg
cm$^{-2}$ s$^{-1}$.  The RMS error in the flux in the same units is
given in Column 4.  Columns 5, 6 and 7 contain the equivalent widths
$W_{\lambda}$ in Angstrom  of the three brightest Balmer lines.  The low
continuum levels do however make the determination of equivalent
widths uncertain.  In general these could be determined only to an
accuracy of $\sim$5 Angstrom  at best.  Any correction for absorption by
the underlying stellar population ($\sim 2$ Angstrom  for normal galaxies
[McCall et al.\ 1985]) will thus be negligible compared to the uncertainties
in the widths themselves.

Column 8 contains the reddening coefficient $c$, as defined in e.g.
Osterbrock (1989).  $c$ is related to $E(B-V)$ by $E(B-V)=0.78c$.
Column 9 contains the error in $c$.  $c$ was determined from the ratio
between the H$\alpha$ and H$\beta$ fluxes and by assuming that the H{\sc ii}
regions were Case B H{\sc ii} regions with a temperature of 10,000 K (see
Osterbrock 1989).  We will show in Sect.~4.1 that this is a justified
assumption. The interstellar extinction curve by Savage \& Mathis
(1979) was used.  The calibration uncertainty of 10\% mentioned above
results in an uncertainty of 0.2 in the reddening coefficient.  In a
few cases negative reddenings were found.  Modestly negative
reddenings can usually be explained by assuming a higher temperature
for the nebula combined with zero reddening.  A few very negative
reddenings could not be explained satisfactorily.  These are noted
with colons in Table 1.  The spectra with negative reddenings
generally have low H$\alpha$ and H$\beta$ fluxes thus introducing
additional uncertainties.
Negative reddenings
have been set to zero in any further analysis.

\begin{table*}
\begin{minipage}{110 mm}
\caption[]{Miscellaneous Fluxes}
\begin{tabular}{lcrrrrrrrr}
\hline
{  Galaxy}& { \ H{\sc ii}}& {  $W_{\lambda}$} & {  [O {\sc ii}]}& 
{  [O {\sc iii}]}&{  H$\alpha$}&
{  [N {\sc ii}]}&{  [S {\sc ii}]}&{  [S {\sc ii}]}&H$\beta$\\
& &$(H\alpha)$&$\lambda 3727$ &$\lambda 4959$ &$\lambda 6563$ & $\lambda 6583$
&$\lambda 6717$ & $\lambda 6731$ &\\
\hline
F563-V1 &3-1& 16&& 88& 76&&&&$<$29\\
F568-3&2-2& 24&&& 78& 21&&&$<$15\\
&2-3&   7&   &        &   115&   33&   39&        &$<$45\\
        &      2-4&   29&  &        &   316&   69&   85& &$<$40\\
F568-V1 &      3-2&   11& 467&        &   377&   115&   122&   145&$<$58\\
        &      3-3&   24& 371&   342&   275&        &        &    68&$<$54\\
        &      3-4&   157& 328&        &   188&   21&        &        &$<$45\\
U1230   &      1-2&   91&   30&        &   317&    29&   32&   39&$<$68\\
        &      1-3&   89&        &        &   304&   32&   46&   33&$<$61\\
        &      3-5&   86&        &        &   72&   55&     109& &$<$57\\
U5005   &      3-1&   40&        &        &   174&   20&   52&   40&$<$28\\
        &      3-4&   33&   306&   178&   341&   42&   94&   62&$<$49\\
U5999   &      4-2&   50&   568&        &   621&   60&     220&   96&$<$71\\
        &      4-3&   44&   29&        &     529&    26&   40&   61&$<$12\\
U6614   &      1-1&   18&        &        &   213&   118&        & &$<$60\\
        &      1-2&   34&        &        &     243&    76&        & &$<$60\\
        &      2-1&   39&        &        &   251&   97&   32&  &$<$22\\
        &      2-2&   349&        &        &   57&   &        &  &$<$14\\
        &      4-a&   17&   776&        &    1294&   552&        &  &$<$124\\
        &      4-1&   27&        &        &    1307&   576&        & &$<$160\\
        &      4-2&   10&        &        &   187&     122&        & &$<$106\\
\hline
\end{tabular}
\end{minipage}
\end{table*}

The errors in the strong lines are shot noise limited, while the
errors in the faint lines are dominated by the read-out noise of the
CCD.  An estimate of the uncertainty in the flux of the
H$\beta$ lines was made by adding in quadrature the shot noise in the
line, the shot noise in the continuum and the read-out noise of the
CCD, which is $\sim 1\cdot 10^{-17}$ erg s$^{-1}$ cm$^{-2}$ for the
red images, and $\sim 5\cdot 10^{-18}$ erg s$^{-1}$ cm$^{-2}$ for the
blue ones.  To ascertain that these were realistic estimates of the
uncertainties, we also determined the uncertainties in the H$\beta$
flux by making two additional measurements of the line-flux, with the
continuum levels systematically offset by $+1\sigma$ and $-1\sigma$ of
the continuum.  Both uncertainties agreed within a factor of two with
each other, where for all but the strongest lines the shot noise value
was slightly smaller than the offset-continuum value.  The shot noise
values are given in Column 4 of Table 1.

Table 2 contains the ratios of the fluxes of the other bright emission
lines with respect to the H$\beta$ flux.  If additional faint lines
were measured in the spectrum the H{\sc ii} region number is marked with a
star.  These extra line fluxes are given in the lower panel of Table
2.

In a few spectra the H$\beta$ line was not detected.  Fluxes of the
lines that were detected in these spectra are given in Table 3.  Also
given in Table 3 is a $1\sigma$ (of the continuum) upper limit of the
H$\beta$ flux. All flux values are expressed in units of $10^{-18}$
erg s$^{-1}$ cm$^{-2}$.

\section{Determining the oxygen abundances}

The standard method to determine the oxygen abundance in a direct way
is described by Osterbrock (1989) and involves determining the
electron temperature of the gas in the H{\sc ii} region by measuring the
ratio of intensities of the strong [O~{\sc iii}]$\lambda\lambda 4959,5007$
lines and the faint [O~{\sc iii}]$\lambda 4363$ line.  However, as can be
seen in Table 2, this line was detected only with certainty in two of
the best spectra (F563-1[2-3] and U5005[2-2]).  We will determine the
oxygen abundance of these two H{\sc ii} regions using the standard method in
Sect.~4.1. For the other spectra the standard method does not suffice
and in Sect.~4.2 we will use an empirical oxygen abundance indicator
to estimate the oxygen abundances.

\subsection{The standard method}

We used the {\sc fivel} program (De Robertis et al. 1987) as
implemented in the {\sc iraf} package to iteratively solve for the
electron temperature for the O$^{++}$ region and the electron density.
We refer to Table 4 for a listing of the relevant values. To determine
the temperature the [O~{\sc iii}] $\lambda\lambda 4959,5009$/ [O~{\sc
iii}] $\lambda 4363$ ratio was used; for the density the [S~{\sc ii}]
$\lambda 6717$/ [S~{\sc ii}] $\lambda 6731$ ratio. The values thus
derived are consistent with the standard values for a low-density
nebula of $T_e = 10000$ K and $n_e = 100$ cm$^{-3}$.

The temperature of the O$^+$ region was calculated using the
relationship given in Campbell et al. (1986):
$$T_e(O^+) = T_e(O^{++}) - 0.3[T_e(O^{++})-1.0], $$
with $T_e$ expressed in units of 10$^4$ K.  The {\sc fivel} program
was used to calculate the O$^{++}$/H$^{+}$ and the O$^{+}$/H$^{+}$
ratios, given the values of the electron density and the fluxes in the
[O~{\sc ii}]$\lambda 3727$ line (which is blended with [O~{\sc ii}]$\lambda 3729$)
and the [O~{\sc iii}]$\lambda 5007$ line.
The total oxygen abundance is then given by O/H $\equiv$
O$^{+}$/H$^{+}$ + O$^{++}$/H$^{+}$.

We thus derive values of log(O/H) $= -3.81$ for F563-1 region 2-3,
and log(O/H) $= -3.83$ for U5005 region 2-2. This should be compared with the
empirical values derived for these regions (in the next section) of
$-3.99$ and $-3.93$ respectively. The empirical values are consistent
with the true values within 0.2 dex, which, as we will show in the
next section, is the typical uncertainty of the empirical method.

\begin{table}
\begin{minipage}{81 mm}
\caption[]{Oxygen: standard method}
\begin{tabular}{lrr}
\hline
 & F563-1:2-3 & U5005:2-2 \\
\hline
[O {\sc iii}] ratio    & 134.76  &  135.24 \\
$\rm{[S {\sc ii}]}$ ratio     & 1.276   &  1.176  \\

T$_e$[O {\sc iii}] (K)    & 11525   &  11509  \\
T$_e$[O {\sc ii}] (K)     & 11068   &  11057  \\
$n_e$ (cm$^{-3}$)           & 165     &  291    \\

log O$^{++}$/H$^+$   & $-3.92$ & $-4.10$ \\
log O$^{+}$/H$^+$    & $-4.50$ & $-4.17$ \\
 
log O/H              & $-3.81$ & $-3.83$ \\
log (O/H)$_{\rm emp}$& $-3.99$ & $-3.93$  \\
\hline
\end{tabular}
\end{minipage}
\end{table}

\subsection{The strong line method}

For spectra where the faint [O {\sc iii}]$\lambda 4363$ line was not detected
the empirical oxygen abundance indicating line ratio R$_{23}$ was
used. This ratio is defined as $${\hbox{R}}_{23} \equiv {{\hbox{[O
      {\sc ii}]}\lambda 3727 + \hbox{[O {\sc iii}]}\lambda \lambda
    4959,5007}\over{\hbox{H}\beta}}.$$ Pagel  et al.\ (1979) were
the first to investigate the behaviour of this line ratio. 
The main problem with this empirical method is the calibration:
reliable measurements of the faint [O {\sc iii}]$\lambda 4363$ line have to
be available.  For our purposes, we will use the calibration and
models presented in McGaugh (1991).

For H{\sc ii} regions with low abundances \r23 does not depend solely on
oxygen abundance, but also on the degree of ionization of an H{\sc ii}
region.  This degree of ionization is given by the volume averaged
nebular ionization parameter $\langle U \rangle$, which is defined by
$$U(R) = {{Q}\over{4\pi c N R^2}}$$ where $Q$ is the ionizing
luminosity, $N$ the gas density, and $R$ the distance from the
ionizing source.  $$\langle U \rangle = {{\int^{R_s}_0
    U(R)dV}\over{\int_0^{R_s} dV}}={{3Q}\over{4\pi cNR_s^2}},$$ for
constant density models, where $R_s$ is the Str\" omgren radius.
$\langle U \rangle$ is then essentially the ratio of ionizing photon
density to mass density.

In order to distinguish between the different degrees of ionization of
the nebulae the ionization sensitive line ratio \o32, defined as
$$\hbox{O}_{32} \equiv {{ \hbox{[O {\sc iii}]}\lambda \lambda
    4959,5007}\over{\hbox{[O {\sc ii}]}\lambda 3727}}$$ is introduced.

This results in a model surface (Fig. 2) defined by contour lines with
constant oxygen abundance and ionization parameter, as a function of
\r23 and \o32.  The exact position of the grid in the \r23--\o32-plane
depends slightly on the assumed upper mass cut-off $M_u$ of the
IMF. The model presented here assumes $M_u = 60 M_{\odot}$.

The oxygen abundance on the upper branch (drawn with
solid lines) is relatively independent of the degree of ionization,
while on the lower branch (dashed lines) the oxygen abundance cannot
be determined reliably without taking \o32 into account.

It is clear that most of this model surface is degenerate: once \r23
and \o32 are determined, there is still a choice between the
low-abundance lower branch and the high-abundance upper branch.
Especially for low \r23 the difference between the two possibilities
can be quite large.

\begin{figure*}
\epsfxsize=0.95\hsize
\centerline{\epsfbox{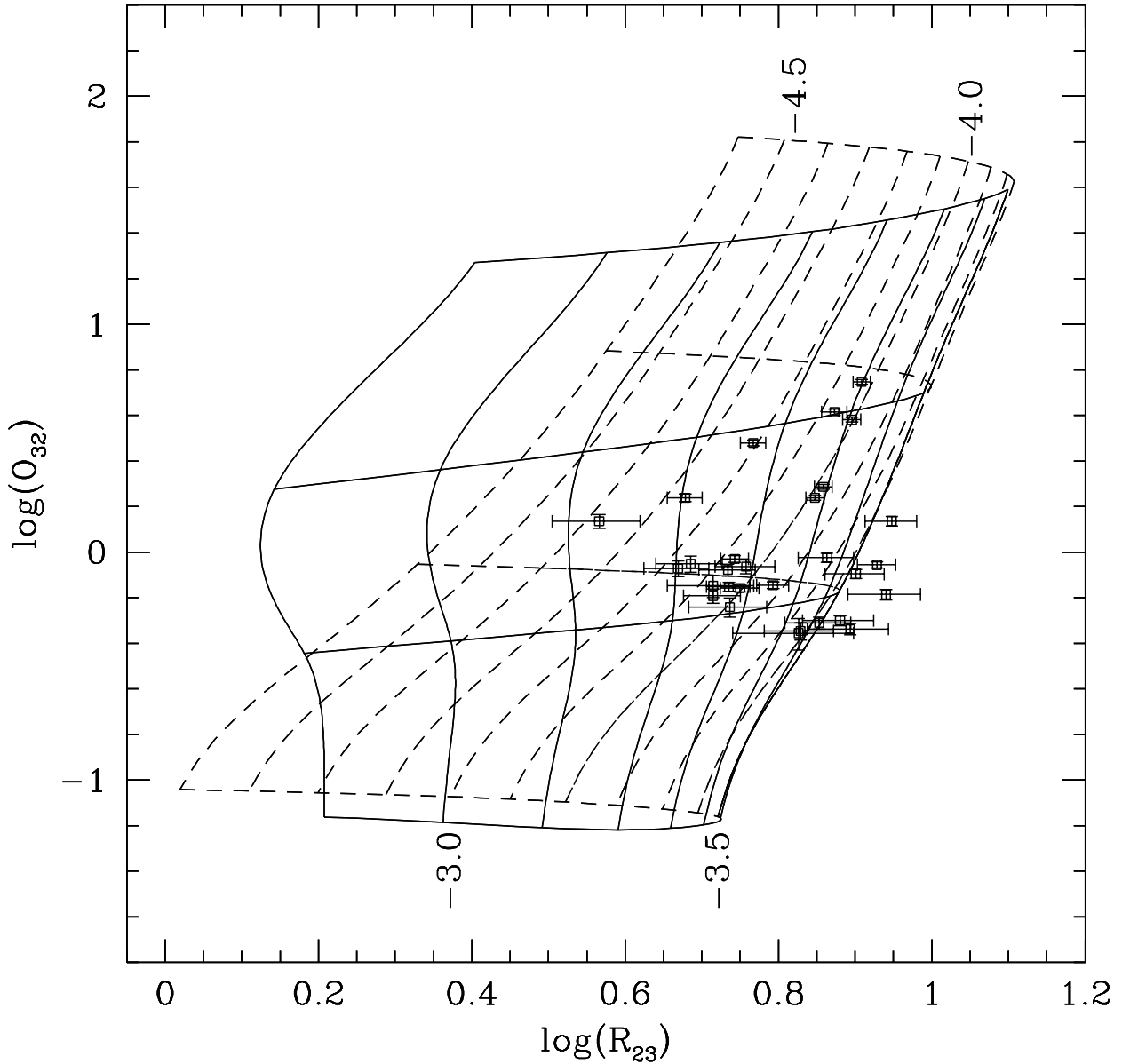}}
\caption{Model grid of McGaugh (1991). The solid lines represent
  the upper branch; the dashed lines the lower branch. The vertical
  lines are lines of constant abundance; the horizontal lines of
  constant \<U>, with log\<U> = --4 at the bottom, and log\<U>= --1 at
  the top. The squares represent the values found for H{\sc ii} regions in
  our sample.}
\end{figure*}

The [N {\sc ii}]$\lambda 6583$ line can be used to find out if a point lies
on the upper or lower branch. The ratio [N~{\sc ii}]/[O~{\sc ii}] varies
monotonically with abundance, and is not very sensitive to $\langle U
\rangle$ (see McCall et al.\ 1985).  H{\sc ii} regions with $\log(\hbox{[N
  {\sc ii}]}/\hbox{[O {\sc ii}]}) > -1$ are on the upper branch.  Most HSB spiral
H{\sc ii} regions are found there (McCall et al.\ 1985).  If $\log(\hbox{[N
  {\sc ii}]}/\hbox{[O {\sc ii}]}) < -1$ the H{\sc ii} regions are on the lower branch.

The model surface has a turnover region or fold at $-3.9<\ $\logoh$\
<-3.4$.  In this region a large change of oxygen abundance
corresponds to only a small interval in \r23.  For H{\sc ii} regions
occupying that part of the diagram an at least partly artificial
crowding will occur around $\log(\hbox{O}/\hbox{H}) = -3.6$.

McGaugh (1991) showed that on the upper branch (solid lines; \logoh\
$>-3.4$ or $Z > 0.5 Z_{\odot}$) the uncertainty in the calibration of
\logoh is $\sim 0.1$ dex, and the uncertainty in the calibration of
$\langle U \rangle$ is $\sim 0.15$ dex.  On the lower branch (dashed
lines; $ \log(\hbox{O}/\hbox{H}) <-3.9$ or $Z<0.15Z_{\odot}$) the
calibration uncertainties are smaller, as the oxygen lines dominate
the cooling process.  The uncertainty in calibrating \logoh is $\sim
0.05$ dex, and that in $\langle U \rangle$ is $\sim 0.1$ dex. To get
an idea of the true uncertainties in the abundance determinations,
these calibration uncertainties should thus be added to the
observational uncertainties (see Sect.~5). For a more complete
discussion of the calibration see McGaugh (1991).
\bigskip

\section{Oxygen Abundances}

Figure 2 shows the reddening corrected positions of the measured H{\sc
ii} regions in the diagram.  In order to avoid spuriously large values
of, especially, $R_{23}$ we have omitted all H{\sc ii} regions where
the peak value of the measured H$\beta$ line was less than 3$\sigma$
above the noise of the continuum. These regions are denoted in Table 2
by having their identification number (column 2) between brackets.
The errorbars are determined by taking into account the uncertainties
in the fluxes of the relevant lines using the continuum-offset
procedure (see above).  Note that the calibration uncertainties of the
model (Sect.~4) are much larger than the formal errors in \r23 and
\o32.

All points are consistent with the model computed assuming an upper
mass cut-off in the Initial Mass Function of $M_u = 60 M_{\odot}$
presented here. We therefore find no strong evidence for the existence
of super-massive stars in the H{\sc ii} regions.

Figure 3 shows the reddening corrected [N {\sc ii}]/[O {\sc ii}] ratio
as a function of \r23 for our H{\sc ii} regions, along with those
given in McCall (1985).  We see that most of our H{\sc ii} regions
have $\log(\hbox{[N {\sc ii}]}/\hbox{[O~{\sc ii}]} < -1$, which
implies that they are on the low-abundance lower branch of the model
grid.

\begin{figure}
\epsfxsize=\hsize \centerline{\hbox{\epsfbox[50 60 775
585]{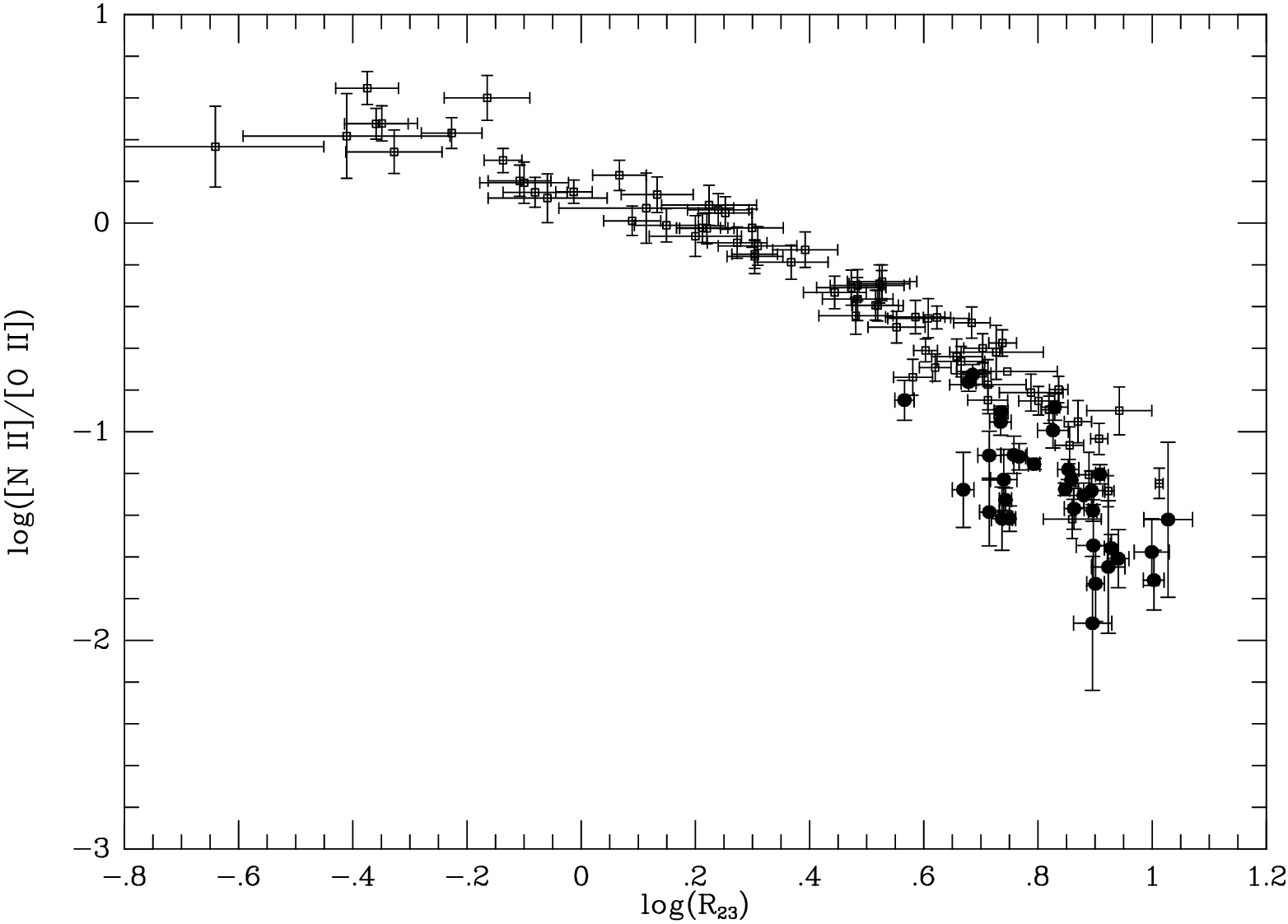}}} \caption{Reddening corrected [N{\sc ii}]/[O{\sc
ii}] ratios as function of \r23 parameter. The filled circles
represent our sample; the small squares the data of McCall et al
(1985).}
\end{figure}

Oxygen abundances and ionization parameters are presented in Table 5,
for those spectra where [N {\sc ii}]/[O {\sc ii}] could be determined.
The errors were determined by projecting the continuum-offset
uncertainties in $R_{23}$ and $O_{32}$ on the \logoh --\<U> -grid, and
adding these in quadrature to the calibration uncertainties in \logoh\
and \<U>\ at these positions. In most cases the errors are dominated by the calibration uncertainties.

Table 6 presents the results for those spectra where [N {\sc ii}]/[O
{\sc ii}] was not measured, and the abundance determination was
ambigious. Both upper and lower-branch values are given.
\begin{center}
\begin{table*}
\begin{minipage}{80 mm}
\caption[]{Oxygen Abundances}
\begin{tabular}{lrrrrr}
\hline
{Galaxy}& {H{\sc ii}}&log O/H &$\sigma_{O/H}$&$\log \langle U \rangle$&
$\sigma_{\langle U \rangle}$\\
\hline
F561-1   &      1-1& --3.28&0.12& --2.69&0.16\\
         &      1-2& --4.03&0.06& --2.89&0.10\\
F563-1   &      2-1& --4.01&0.08& --3.06&0.11\\
         &      2-3& --3.99&0.06& --2.05&0.10\\
         &      4-1& --3.96&0.11& --3.10&0.11\\
         &      4-2& --3.95&0.06& --2.22&0.10\\
F563-V2  &      2-1& --3.25&0.10& --2.34&0.15\\
         &      2-2& --4.17&0.06& --2.36&0.10\\
         &      2-3& --3.80&0.23& --2.85&0.20\\
         &      4-1& --3.60&0.24& --2.86&0.20\\
         &      4-2& --3.60&0.28& --3.12&0.20\\
F568-3   &      2-1& --3.32&0.11& --2.74&0.15\\
F571-5   &      3-2& --3.93&0.06& --2.53&0.10\\
U1230    &      1-1& --3.16&0.11& --2.43&0.15\\
U5005    &      2-2& --3.93&0.06& --2.58&0.10\\
         &      2-4& --3.60&0.23& --3.08&0.20\\
         &      2-5& --3.60&0.27& --3.09&0.23\\
         &      3-2& --4.14&0.09& --2.95&0.11\\
         &      3-3& --3.99&0.11& --2.92&0.11\\
         &      3-5& --4.03&0.11& --3.01&0.11\\
         &      4-1& --3.89&0.07& --2.99&0.10\\
U5999    &      3-1& --3.60&0.25& --2.99&0.20\\
         &      3-2& --3.60&0.27& --3.14&0.20\\
         &      3-3& --3.60&0.25& --2.91&0.20\\
         &      3-4& --3.97&0.07& --3.01&0.10\\
         &      4-1& --3.60&0.30& --3.16&0.20\\
         &      4-4& --3.33&0.10& --2.83&0.15\\
\hline
\end{tabular}
\end{minipage}
\end{table*}
\end{center}
\begin{table*}
\begin{minipage}{100 mm}
\begin{center}
\caption[]{Ambiguous Abundances}
\begin{tabular}{lrrrrrrrrr}
\hline
{Galaxy}& {H{\sc ii}}&log O/H &$\sigma_{O/H}$&$\log \langle U \rangle$&
$\sigma_{\langle U \rangle}$& log O/H &$\sigma_{O/H}$&$\log \langle U \rangle$&
$\sigma_{\langle U \rangle}$\\
\hline
F563-V1   &     3-2& -3.30&0.16& -2.57&0.15& -3.60&0.23& -2.66&0.20\\
F571-5    &     3-1& -3.39&0.11& -1.98&0.15& -4.02&0.05& -2.20&0.10\\
\hline
\end{tabular}
\end{center}
\end{minipage}
\end{table*}

Figure 4a shows histograms with the distribution of oxygen abundances.
The top panel gives the distribution of abundances of H{\sc ii} regions from
our sample, the middle panel shows the distribution of oxygen
abundances of H{\sc ii} regions from the sample of McGaugh (1994) (his Table
3).  The bottom panel shows the distribution of both samples combined.
Both samples have a few galaxies in common, but the number of H{\sc ii}
regions that are in both samples is a few at most and any overlap will
not influence the eventual results.

\begin{figure*}
\begin{center}
\hbox{\epsfxsize=0.45\hsize\epsfbox[40 70 570 740]{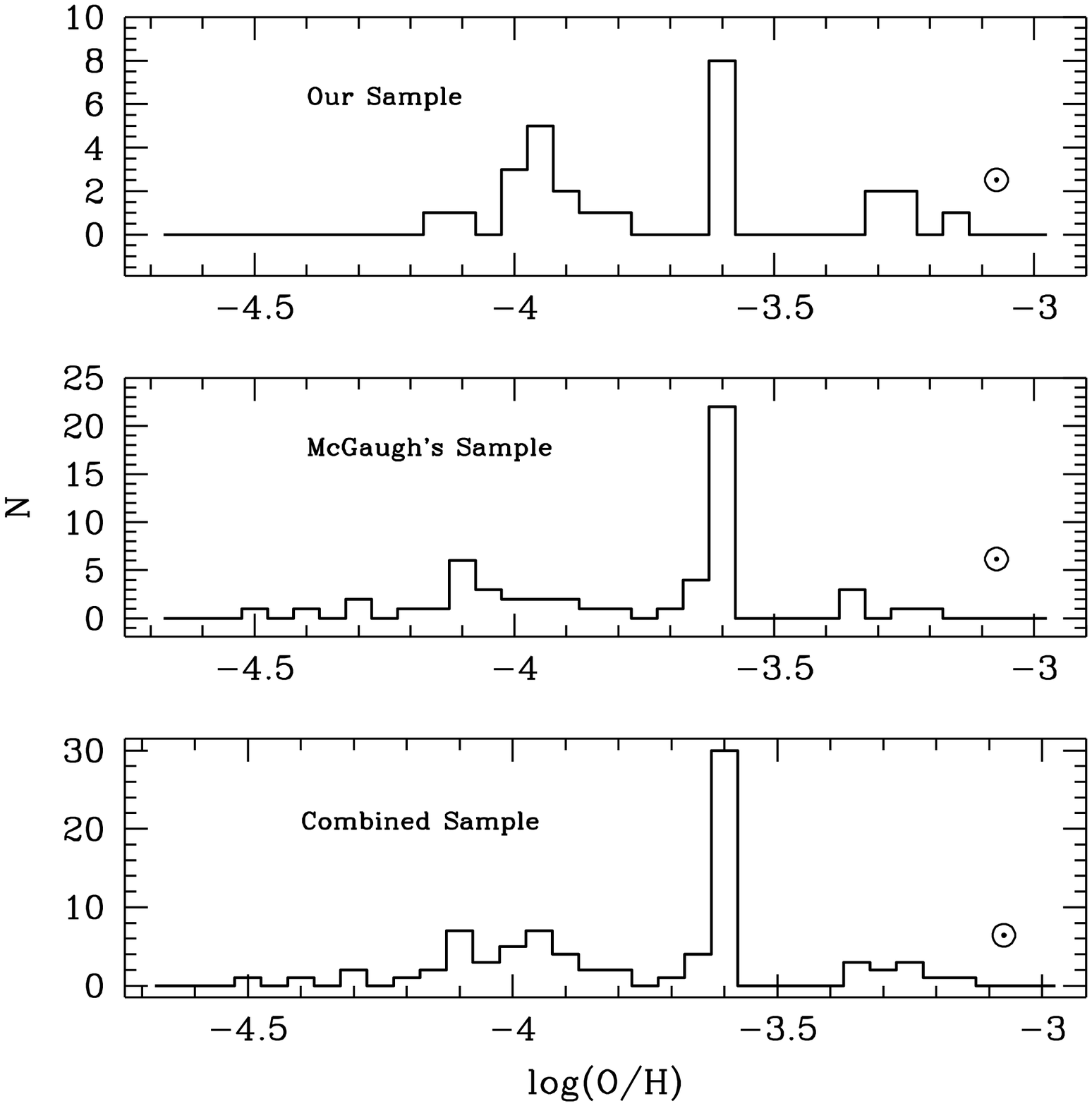}\hspace{10mm}\epsfxsize=0.45\hsize\epsfbox[40 70 570 740]{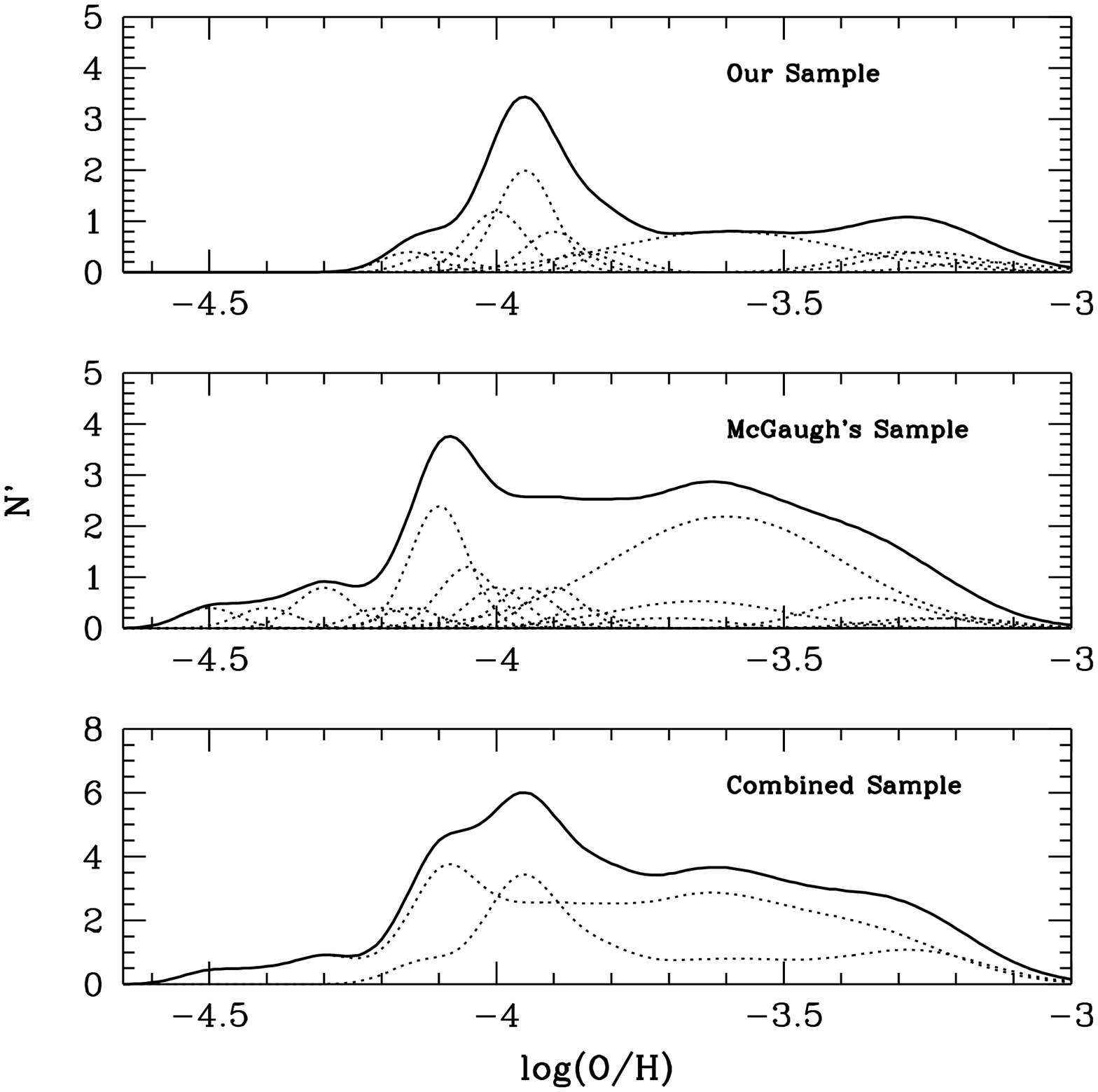}}
\caption{Left: Histograms of oxygen abundances of H{\sc ii} regions in LSB
  galaxies. Right:Histograms of oxygen abundances of H{\sc ii} regions in
  LSB galaxies.  The histogram bars have been replaced by gaussians
  with error dependant widths. See text for more details. }
\end{center}
\end{figure*}

The peak in the histograms at \logoh = --3.6 in Fig.\ 4a is at least
partly artificial.  This is due to the fold at \logoh = --3.6 in the
model grid of Fig.\ 2. We have attempted to correct for this in Fig.\ 
4b.  Here each of the histogram bins has been replaced by a gaussian
with a width $\sigma$ (= 0.43 FWHM) equal to the uncertainty of the
model grid at that abundance (e.g., the gaussian at \logoh = --3.6 has
$\sigma = 0.2$). The peak value was determined keeping the area under
the gaussian equal to that of the corresponding histogram bar.  As a
result the peak at \logoh = --3.6 has been smeared out, and the
distribution is almost flat, with maybe a slight peak at \logoh $\sim
-4$.  This peak might be the result of our selection method.  Low
abundance regions tend to be more ionized, and therefore brighter
(e.g., Campbell 1988; Dopita \& Evans 1986).  Regardless of whether
this effect is present or not, it is clear that within the abundance
range shown by LSB galaxies, there is no preferred value.

With regard to the lowest abundances, Kunth \& Sargent (1985) note
that values of \logoh = --4.3 can already be reached after the first
generation of massive stars. This might explain the cut-off in the
abundance distribution around \logoh = --4.3 in Fig. 4. LSB galaxies
are not primordial objects, but some of them appear to be very
unevolved. Furthermore, to retain such low abundance values at the
present epoch, these galaxies must have been quiescent over their
entire life time, and some of them may not evolved significantly since
their first epoch of star formation.

\section{Reddening and extinction}

The top panel in Fig.\ 5 contains a histogram of the reddenings
towards the H{\sc ii} regions in our sample, corrected for the galactic
contribution (Burstein \& Heiles 1984). The middle panel contains a
histogram for the sample of McGaugh (1994), while the bottom
panel shows both samples combined.  It is clear that the extinction in
H{\sc ii} regions is low with $\langle A_V \rangle = 3\langle E(B-V) \rangle
\simeq 0.5$ mag. For normal HSB galaxies extinction values are usually
found to be up to a factor 6 larger (see e.g.\ Fig.~7 in Zaritsky et al.
1994).  The conclusion that can be drawn from
this, is that the reddening towards these H{\sc ii} regions is low, but at
some sites some dust is present.

In the lowest abundance H{\sc ii} regions star formation is apparently able
to proceed without much dust. These low reddenings also confirm that
dust is not the cause of the low surface brightness of LSB galaxies.

\section{Gradients}

For galaxies F563-1, UGC 5005 and UGC 5999 we have enough measurements
over a large enough radial range that we may attempt to investigate
possible radial oxygen abundance gradients. We have plotted in Fig. 6 the
abundances of the H{\sc ii} regions in these galaxies versus the deprojected
radial distance, expressed in scale lengths.
There is no clear trend of abundance with radius. Rather the oxygen
abundance seems to be constant at $-3.8 \pm 0.5$.

\begin{figure}
\epsfxsize=0.9\hsize
\epsfbox[7 62 600 731]{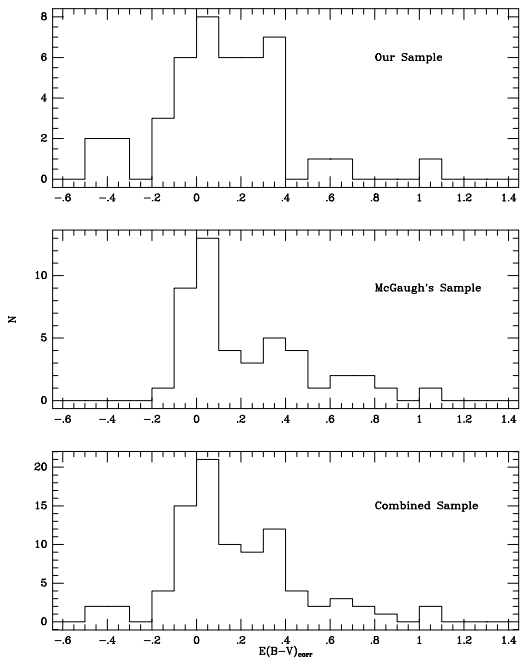}
\caption{Histograms of reddenings towards H{\sc ii} regions in LSB galaxies.}
\end{figure}

\begin{figure}
\epsfxsize=\hsize
\epsfbox{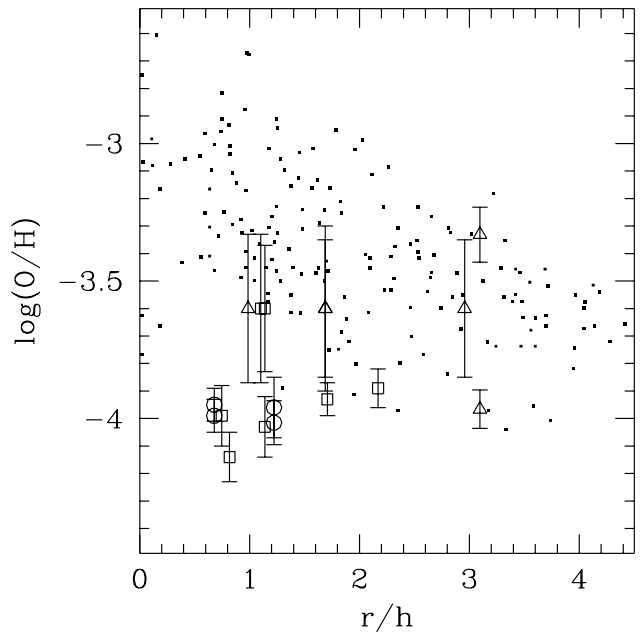}
\caption{Oxygen abundances in H{\sc ii} regions of 3 LSB galaxies. The true 
  radial distance to the center (corrected for inclination effects) is
  expressed in optical scale lengths. The circles denote galaxy
  F563-1; the squares denote UGC 5005 and the triangles UGC 5999. No
  strong trend is seen. HSB late-type galaxies show a gradient of
  $\sim 1$ dex in (O/H) over the radial range shown here.}
\end{figure}

To compare with other galaxies of similar Hubble type, we show in Fig.
6 the abundances of H{\sc ii} regions in 9 late-type Scd-Irr galaxies, taken
from VCE (their Fig. 2). We have converted their half light radii
$R_d$ to disk scale lengths using the relation $R_d = 1.678 h$, under
the implicit assumption that the light distribution of their sample
galaxies can be described by an exponential disk.  As we are dealing
with very late-type galaxies this assumption is most likely justified.

The VCE galaxies show a gradient of $\sim 1$ dex over the range in
radius sampled by the LSB galaxies. The LSB H{\sc ii} regions clearly do not
follow this trend.  The oxygen abundances in the LSB galaxies are
lower at all radii, except in the very outer parts where the abundance
values of the HSB and LSB galaxies appear to converge.  The lack of
abundance gradients in LSB galaxies may indicate that the picture of
galaxies evolving from the inside out may not apply to LSB galaxies.
It is usually assumed that the abundance and colour gradients found in
galaxies indicate that the outer parts of galaxies are less evolved
than the inner parts.  The colours and abundances in LSB galaxies are
comparable with those in the outer (most unevolved) parts of HSB
galaxies.  The absence of a radial abundance trend in the LSB galaxies
suggests that these may have evolved at the same rate over their
entire disk. Indeed, a spatially and temporally sporadic star
formation rate, as is derived for LSB galaxies (e.g. Gerritsen \& de
Blok 1997), would not give rise to an abundance gradient.

Alternative explanations for the lack of a gradient may be the infall
of metal-poor gas from high above the planes of these galaxies.
However, because of the low star formation activity in LSB galaxies it
is not likely that large amounts of gas can have been blown out in the
past.  If gas infall is a major factor, this must be ``primordial''
gas, left over from when the LSB galaxies were formed. 

A second alternative is that the disks of LSB galaxies are still
settling in their final configuration. Gas from larger radii is slowly
diffusing inward, causing density enhancements where conditions for
star formation may be favourable. This is qualitatively consistent
with the finding that most of the star forming regions in LSB galaxies
are found towards the outer radii of the stellar disk.  If star
formation in LSB galaxies needs to be stimulated by external
conditions (like infall), and is not self-propagating, this building
up of the disk would not give rise to an abundance gradient. If every
radius would go through one cycle of star formation before fading,
after which star formation would move on to larger radii, this would
cause a colour gradient (as is observed) but no abundance gradient.
Note that though this evolution ``from the inside out'' is different
from the standard picture mentioned above. In that picture each radius
continues evolving, with the inner radii going through more star
formation cycles than the other radii.
Clearly, the scenario described here is only a crude ad-hoc attempt to
explain the lack of gradient. It would have to be tested by modelling,
and supported by more abundance data.  

In this respect the discussion presented in Edmunds \& Roy (1993) is
of interest.  They show that steep abundance gradients in gas-rich
disk galaxies seem to require the presence of unbarred spiral
structure. The abundance gradients disappear at the same absolute
magnitude that spiral structure ceases ($M_B \sim -17$) and are
considerably shallower in galaxies with a strong bar.

The late-type LSB galaxies in general show only a faint spiral
structure, and have $M_B \sim -17.5$. Their feeble spiral structure
may thus be related with the lack of abundance gradients. Edmunds \&
Roy (1993) offer two possible explanations. The variation with radius
of the frequencey with which interstellar material passes through a
spiral pattern may result in a declining star formation rate with
radius. This would set up an abundance gradient. At lower absolute
magnitudes star formation will still continue, but the spiral pattern
would be absent. This would result in a star formation rate which is
much more uniform across the disk than in strong
spirals. Aternatively, if non-circular motions are relatively more
important the resulting mixing may also homogenize the chemical
composition of the interstellar medium.

Although we realize that it is easy to over-interpret data of such a
limited dynamic range, we conclude that the lack of abundance
gradients confirms the evolutionary picture of quietly evolving LSB
galaxies with only local processes regulating their evolution.

\section{Conclusions}

The oxygen abundances in 64 H{\sc ii} regions in 12 LSB galaxies have
been measured.  Oxygen abundances are low. No region with solar
abundance has been found, and most have oxygen abundances that are
$\sim 0.5$ to 0.1 solar.  No strong radial oxygen abundance gradients
are found. The abundance seems to be constant, rather, as a function
of radius, supporting the picture of quiescently and sporadically
evolving LSB galaxies.

\begin{acknowledgements}
We thank the referee Dr.~Smartt for his constructive comments which have helped improve the presentation of these data.

The William Herschel Telescope is operated on the island of La Palma by
the Isaac Newton Group in the Spanish Observatorio del Roque
de los Muchachos of the Instituto de Astrof\i\' sica de Canarias.

\end{acknowledgements}

\end{document}